\documentclass[11pt]{article}

\usepackage{savesym}
\usepackage{amsmath}
\usepackage{booktabs}
\usepackage{pdflscape}
\usepackage{float}
\usepackage{threeparttable}
\usepackage{multirow}
\usepackage[table]{xcolor}
\usepackage{setspace}
\savesymbol{tnote}

\usepackage{lineno}

\usepackage{caption}
\usepackage{ctable}
\restoresymbol{try}{tnote}
\usepackage{amssymb}
\usepackage{graphicx}
\usepackage{anysize}
\marginsize{3.8cm}{3.2cm}{2cm}{3.9cm}
\makeatletter
\renewcommand\@biblabel[1]{}
\renewenvironment{thebibliography}[1]
     {\section*{\refname}%
      \@mkboth{\MakeUppercase\refname}{\MakeUppercase\refname}%
      \list{}%
           {\leftmargin0pt
            \@openbib@code
            \usecounter{enumiv}}%
      \sloppy
      \clubpenalty4000
      \@clubpenalty \clubpenalty
      \widowpenalty4000%
      \sfcode`\.\@m}
     {\def\@noitemerr
       {\@latex@warning{Empty `thebibliography' environment}}%
      \endlist}
\makeatother

\title{Epidemic Modelling of Bovine Tuberculosis in Cattle Herds and Badgers in Ireland}

\begin{document}

%\modulolinenumbers[1]
%\linenumbers
%\date{}
\date{\vspace{-5ex}}
\maketitle

\begin{center}
\begin{tabular}{cc}
L.M. White* & $^\dag$G.E. Kelly\\
Visiting Nurse Service of New York& School of Mathematics and Statistics\\
(VNSNY)& University College Dublin\\
& Ireland
\end{tabular}
\end{center}
*Work undertaken as an M.Sc. student at University College Dublin. The paper does not represent the views of VNSNY\\
$\dag$ corresponding author: gabrielle.kelly@ucd.ie
\section*{Abstract}
Bovine tuberculosis, a disease that affects cattle and badgers in Ireland, was studied via stochastic epidemic modeling using incidence data from the Four Area Project (Griffin et al., 2005). The Four Area Project was a large scale field trial conducted in four diverse farming regions of Ireland over a five-year period (1997-2002) to evaluate the impact of badger culling on bovine tuberculosis incidence in cattle herds.

Based on the comparison of several models, the model with no between-herd transmission and badger-to-herd transmission proportional to the total number of infected badgers culled was best supported by the data.

Detailed model validation was conducted via model prediction, identifiability checks and sensitivity analysis.

The results suggest that badger-to-cattle transmission is of more importance than between-herd transmission and that if there was no badger-to-herd transmission, levels of bovine tuberculosis in cattle herds in Ireland could decrease considerably.\\

Keywords: Mycobacterium bovis, Tuberculosis, Epidemic model, Ireland.

\section{Introduction}
A bovine tuberculosis (bTB) (causative agents are any of the disease-causing mycobacterial species within the M. tuberculosis-complex) eradication scheme was
initiated in Ireland in 1954. Although initial progress was good, the programme
subsequently stalled. Presently bTB incidence in cattle herds in Ireland is roughly
4\% and approximately 18,500 infected cattle were slaughtered in 2011, with costs
both to the farmer and the exchequer. BTB also infects wild badgers
(\emph{Meles meles}), a protected species under the Wildlife Act 1964, and they have
been implicated in the transmission of the disease to cattle (Griffin et al., 2005, Kelly et al., 2008). It is also possible for cattle to infect badgers
and the relative contribution of the two species to the persistence of the disease
in cattle is difficult to quantify from field experiments.
\par Epidemic models can play a role in this quantification and here we present
a stochastic model of transmission dynamics of bTB in cattle herds and
wild badgers in Ireland. In contrast to deterministic models that describe
average effects, stochastic models contain and produce variability. For example,
the time when a herd becomes infected and the daily rate of infection is largely
unpredictable. A deterministic model might define the rate of infection as
0.25/herds/day but in a stochastic model it may be defined as 0.14-0.40/herds/day,
i.e. daily rates vary. The probability of each of these rates can be modeled to
form a distribution that peaks at 0.25/herds/day. This variability provides a
range of effects i.e. a confidence interval within which the likely course of an
epidemic will probably lie. To paraphrase De Jong (1995), the gain of
such modeling is not the resulting model, but instead the insight into the
population dynamics of infectious agents that is obtained in the process of model
building and model analysis on the one hand, and interpreting experimental and
observational data on the other.
\par Donnelly and Hone (2010) presented epidemic
models for bTB corresponding to areas of the Randomised Badger Culling Trial
(RBCT) in S-W England. However, transmission dynamics between badgers
and cattle herds may be inherently different in Ireland than in Britain and
require separate modeling. There is a substantial growing body of evidence that badgers in Great Britain are
different in many respects to badgers in Ireland including genetic origins, diet and
territoriality (O'Meara et al., 2012; Sleeman et al., 2008), with Irish badgers exhibiting more wide ranging behaviour (Judge et al., 2009; Kelly et al., 2010).
\par In this study we describe epidemiological models for the transmission of bTB both
from herd-to-herd and from badger-to-herd and apply them to data from the
Irish Four Area Project (FAP), the purpose being to determine the relative contribution of the two infectious pathways. Five subsets of the FAP data are analysed separately.
%Estimates of infected herd numbers for different infected badger population sizes are derived from the models together with %basic reproduction number estimates. Detailed sensitivity analyses of the models are conducted to determine which model %parameters explain
%the most variability in disease prevalence. Results are compared to those of Donnelly and Hone (2010) who applied similar %models to the RBCT data.
\section{Material and Methods}
\subsection{Epidemic Models}
We consider extensions of the classical susceptible-infective-removed (SIR)
epidemic model. One extension was formulated by Barlow et al. (1998) to
explain the transmission of bTB between cattle herds and brush-tail possums
in a region of New Zealand. The model described the transition of herds between
three states: $U$ (uninfected and susceptible to infection), $I$ (infected and
thus infectious, but undiagnosed) and $M$ (under movement control (MC), thus not
infectious to other herds).
\par An analogous model was subsequently formulated by Donnelly and Hone (2010) for
a single area with both risk of infection from wildlife and density-dependent
between-herd risk of infection (i.e. a model which assumes that the rate of
contact of one herd with another increases in proportion to the total number of
herds in the population).
\par
\begin{figure}[H]
\centering
\caption{Representation of the transfer of cattle herds between states: U (uninfected), I (Infected) and M (on Movement Control). Parameters are as follows: $k$ is the rate of infection from badgers to cattle per year, $\beta$ is the between-herd transmission coefficient, $c$ is the rate at which infected herds go on movement control per annum and $p$ is the average length of time in years a herd spends on movement control.
}\label{donnmod}
 \includegraphics[width=0.6\textwidth]{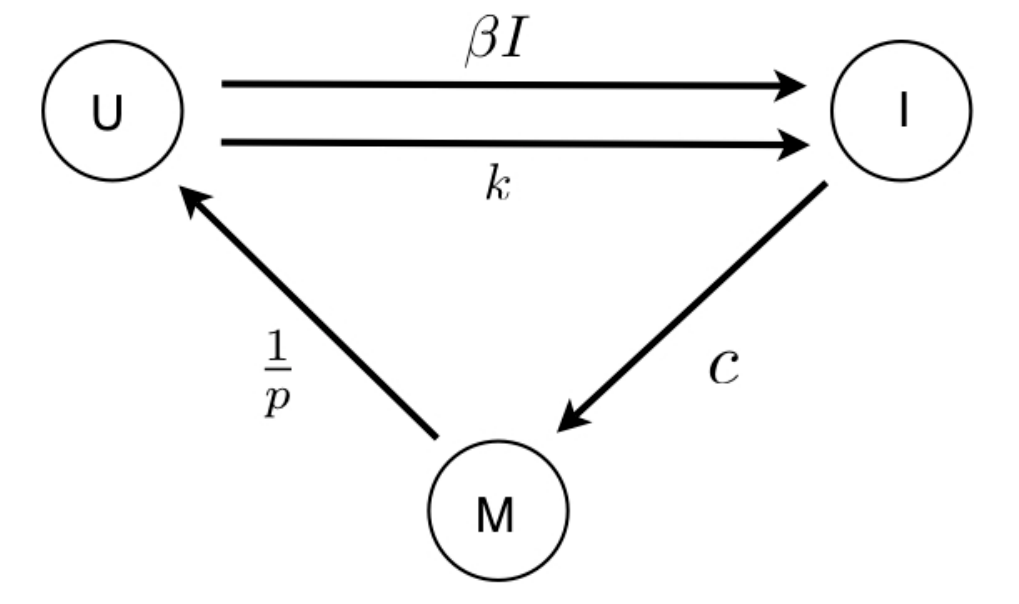}
\end{figure}

 The model is described by the following differential equations:
\begin{align}
\frac{dU}{dt}&=\frac{M}{p}-U(\beta I+k)\label{eq1}\\
\frac{dI}{dt}&=U(\beta I+k)-I c\label{eq2}\\
\frac{dM}{dt}&=Ic-\frac{M}{p}\label{eq3}
\end{align}
with constant population size
\begin{equation}
N=U+I+M
\end{equation}
where $c$ is the overall rate at which infected herds go on MC (year$^{-1}$), $p$
is the average length of time a herd spends on MC in years, hence, $1/p$ is the
rate at which herds come off MC, $k$ is the rate of infection from wildlife to cattle
(year$^{-1}$), and $\beta$ is the between-herd transmission coefficient which
represents the herd-to-herd risk of infection per year. A visual representation of the model is given in Figure~\ref{donnmod}.

Setting Equations~(\ref{eq1}) and (\ref{eq3}) to 0 to find equilibrium values, we
obtain the following quadratic equation for the equilibrium value of $I$, $I^*$.
\begin{equation}
I^{*2}(\beta+pc\beta)+I^*(-N\beta+k+pck+c)-Nk=0
\end{equation}
$I^*$ will later be used in the formulation of a probabilistic model to explain
the proportion of herds with newly detected bTB infection in a year.

In the specific case of no risk of infection from wildlife ($k=0$) and $\beta>0$,
the equilibrium value of $I$ is:
\begin{equation}
I^*=\left(N-\frac{c}{\beta}\right) \left(\frac{1}{1+pc}\right)\label{eqs10s}
\end{equation}
%We arrived at Equation ~(\ref{eqs10s}) by substituting $k=0$ into
%Equations~(\ref{eq1})-(\ref{eq3}) to get $I^*$ when there is no risk of %infection from wildlife.

Similarly, when there is no herd-to-herd transmission of infection ($\beta=0$) and
$k>0$, the equilibrium value of $I$ is:
\begin{equation}
I^*=\frac{N}{1+pc+\frac{c}{k}}\label{mod2eq}
\end{equation}

We also consider a second model which assumes herd-to-herd transmission to be
frequency-dependent i.e. the rate of herd-to-herd transmission is completely
independent of the total number of herds, $N$. The model and associated equilibrium values are described in Donnelly and Hone (2010).
%The model is defined as follows:
%\begin{align}
%\frac{dU}{dt}&=\frac{M}{p}-U\left(\beta \frac{I}{N}+k\right) \label{eq4}\\
%\frac{dI}{dt}&=U\left(\beta \frac{I}{N}+k\right)-I c \label{eq5}\\
%\frac{dM}{dt}&=Ic-\frac{M}{p} \label{eq6}
%\end{align}
%
%Again, we find equilibrium values and obtain the following quadratic equation for
%the equilibrium value of $I$, $I*$:
%\begin{equation}
%I^{*2}\left(\frac{\beta}{N}+pc\frac{\beta}{N}\right)+I^*\left(-\beta+k+pck+c\right)-Nk=0
%\end{equation}
%
%In the specific case of no risk of infection from wildlife ($k=0$) and $\beta>0$
%(using Equations~(\ref{eq4})-(\ref{eq6})), the equilibrium value for $I$ is:
%\begin{equation}
%I^*=N\left(1-\frac{c}{\beta}\right) \left(\frac{1}{1+pc}\right)
%\end{equation}
\par As in Donnelly and Hone (2010), we consider four alternative values of $k$
(the rate of infection from wildlife to cattle per year):
\begin{enumerate}
\item $k=0$: there is no transmission from wildlife.
\item $k= \alpha N_w$: $k$ is proportional to the total number of badgers
    culled in the region in question, $N_w$.
\item $k= \alpha I_w$: $k$ is proportional to the total number of infected
    badgers culled in the region in question, $I_w$.
\item $k= \alpha I_w/N_w$: $k$ is proportional to the proportion of infected
    badgers culled relative to the total number of badgers culled in the
    region in question,
\end{enumerate}
where $\alpha$ is a proportionality constant assumed to be non-negative. The three
types of between-herd transmission (no transmission, density-dependent and frequency-dependent) combined with the four types of transmission from wildlife, gives
eleven different alternatives for $I^*$ (omitting the alternative that has neither
between-herd transmission nor transmission from wildlife, $\beta=k=0$).
\subsection{The Data}\label{data}
Data were taken from the FAP, a large scale field project
undertaken in matched removal and reference areas (each approximately 245 km$^2$) in
four counties in Ireland: Cork, Donegal, Kilkenny and Monaghan over a five-year
period (September 1997-August 2002). The project was carried out to assess the
impact of badger removal on bTB incidence in cattle herds. Badger removal was
proactive in removal areas while minimal culling took place in reference areas.
The badger-removal procedures are described in detail  in the Badger Manual prepared by the Department of Agriculture, Food and Forestry (DAFF, 1996). In summary, badgers were killed be a member of the FAP team after being captured during an 11-night removal operation in which restraints were placed at active setts for 11 nights and were inspected each morning. All euthanased badgers went through gross post-mortem investigation. If evidence of tuberculosis was detected, all affected tissues were sent for histopathological examination and for culture. If no evidence of tuberculosis was detected, bacteriological culture was conducted on multiple tissues, including the lymph nodes, kidney and lung tissue. A badger was diagnosed as positive for tuberculosis if it was positive at histopathological examination and/or culture. The study is described in further detail in Griffin et al. (2005). \par
We consider data on badgers and cattle from the removal areas of the FAP during the study period (1997-2002) and the 5-year 'pre-study period' (September
1992-August 1997). Data on herds were collected routinely by all local District Veterinary Offices and data relating to badger removal and infection status were collected throughout the field trial by FAP staff.
\par Every animal in every herd in Ireland is tested annually for bTB by the Single Intradermal Comparative Tuberculin Test (SICTT) and a herd
is considered positive if any cattle test positive. Herds that test positive are placed under restriction - MC. There is, however, an
incubation period for the disease in cattle; therefore, five subsets of the cattle and
badger data set were considered for analysis, numbered 1-5 as follows:
\begin{enumerate}
\item Data on badgers were taken from the year of the initial badger cull
    (1997/1998) while cattle herd data were taken from the year prior to that.
    These data were chosen because the majority of badgers
were culled in the first year of the study and data on cattle from the year
previous to the initial badger cull had not been affected by the badger
culling and to allow for an incubation period for the disease in cattle.
\item Data on both badgers and cattle were taken from the year of the initial
    badger cull (1997/1998).
\item Data on badgers were obtained by summing over the five years in the study
    period for each area. Data on the number of restricted herds, $B$, were
    also summed over the five years in the study period for each area.
    However, data on the total number of herds, $N$, were not summed as this
    would result in counting the same cattle herds more than once. Therefore,
    for each area, the total number of herds is taken to be the maximum number of
    herds in any one year over the five years.
\item Data on badgers and data on the number of herds restricted were obtained
    by summing over the ten years in the combined study and pre-study period
    in each of the four areas. As above, for each area, the total number of
    herds, $N$, over the ten-year period is taken to be the maximum number of herds
    in any one year over the ten years.
\item Similar to data set 3, except the pre-study period data (1992-1997) were used
    instead of the study period data.
\end{enumerate}

All five data sets are displayed in Table~\ref{datf}. Complete data sets can be found in
Griffin et al. (2005) and Corner et al. (2008). A recent study
suggests a high specificity of between $99.2-99.8$\% for the SICTT in Irish
settings (Clegg et al., 2011). For all data sets above, false positive
misclassification was further minimised by limiting positive infection status to
those SICTT-positive reactors in herds restricted from trading following
disclosure of two or more such animals.

\subsection{Statistical Methods}\label{statmethods}
\subsubsection{The Model}
An epidemic model was employed above to formulate an expression for the
equilibrium value of $I$, the number of infected herds, in terms of the unknown
parameters $\beta$, $\alpha$, $c$ and $p$. We estimated these unknown
parameters by setting up a binomial log-likelihood for the proportion of herds $B$
out of the total number of herds $N$, that experience a bTB herd breakdown and
become restricted in a year. Assuming $B$ is approximately Binomial($N$,
$q=I^*c/N$) at equilibrium, the log-likelihood is:
\begin{equation}
l=\log(L\left (q|N,B \right ))\propto B \log(\left (q \right )) +(N-B)\log(\left (1-q\right
)) \label{bin}
\end{equation}

The model with the value of $I^*c$ closest to $B$ is the 'best' fitting model.
There are four sets of ($B, N$), one set for each of the four counties, hence,  the
log-likelihood is:\\
\begin{equation}
l=\sum_{j=1}^{4} \log(L\left (q_{j}|N_j,B_j \right ))\propto \sum_{j=1}^{4} B_j
\log(\left (q_{j} \right ))+(N_j-B_j)\log(\left (1-q_{j}\right ))\label{sumbin}
\end{equation}
where
\begin{equation}\label{icnj}
q_{j}=I^*_{j}c/N_j
\end{equation}
For each of the eleven alternatives for $I^*$ there is a separate likelihood.

\subsubsection{Parameter Estimation}
\par There are four unknown parameters in each of the eleven likelihoods (via $I^*$),
$\alpha$, $\beta$, $c$ and $p$. These four parameters are assumed to be the same
for each of the four counties. The method of maximum likelihood is used to
estimate $\alpha$ and $\beta$ and associated standard errors (s.e.) for each of
the eleven likelihoods (using Equation~(\ref{sumbin})). Wald's method is then used to construct confidence intervals (C.I.) for the unknown parameters.
\par Empirical values for the other two parameters, $c$ and $p$, which do not vary with the model, are assumed. The parameter c, defined as the rate at which infected herds are detected and put
under MC, incorporates two rates of detection, $1/\mu$, the rate at which infected
herds are detected via the annual SICTT, and $a$, the rate at which infected herds are detected through
slaughterhouse surveillance. As in Cox et al. (2005), we define $\mu$ by:
\begin{equation}
\mu=b\left(\frac{1}{2}+\frac{(1-s)}{s}\right)
\end{equation}
where $s$ is the herd test sensitivity (in per cent) and $b$ is the number of
years between routine herd tests. The first term comes from the assumption that
herds become infected at random times between tests. The second term,
$(1-s)$/$s$, is the number of retests required when the test has less than 100\%
sensitivity, assuming a geometric distribution. Now,
\begin{equation}\label{eq_c}
c=a+\frac{1}{\mu}=a+\frac{2s}{b(s+2(1-s))}
\end{equation}

Values for the parameters $s$, $b$ and $a/c$ (which is the proportion of infected
herds detected through slaughterhouse surveillance) are obtained from the
literature. Frankena  et al. (2007) estimated that in recent years between 27\% and 46\% of
all new herd breakdowns in any year have been detected by slaughterhouse
surveillance in Ireland. Taking the mid-point we approximate $a/c$ by 36.5\%.
However, other values of $a/c$ between 27\% and 46\% were also considered. The
parameter $b$ is equal to 1 since all herds in Ireland are subject to annual
routine herd testing.

Clegg et al. (2011) suggest a sensitivity of $52.9\%-60.8$\% for the SICTT and based
on this the value for the sensitivity, $s$, is taken as the mid-point of this
interval, 56.85\%.

Letting $a=36.5c$, $b=1$ and $s=0.5685$ we solve for $c$ in Equation~(\ref{eq_c})
to obtain $c=1.25$. Since $c$ is the rate at which infected herds are detected, 1/$c$ (0.8 years) is the average length of time in years that a herd is infected before it is detected.
\par We assume that $p$, which is the average length of time a herd spends on MC, is
equal to 0.5 years. When a herd is put on MC it must pass two consecutive SICTT
tests before it is taken off MC and the length of time between retests is
approximately 60 days. Hence, the minimum length of time a herd spends on MC is
approximately 120 days (0.329 of a year) if it passes its first two consecutive
tests. It has been noted that the average restriction period is not very much
longer than 120 days (M. Good, Department of Agriculture, Food and the Marine, personal communication), while in the study of Donnelly and Hone (2010) a value of $p=0.7$ was used. Thus we have let $p$=0.5 a point midway between these two values.

\subsubsection{Measuring Goodness-of-Fit}
\par The relative support for each of the eleven models, $i=1,\ldots,11$, (for the
eleven alternative $I^*$)  was calculated using Akaike weights (Burnham and Anderson, 2002). The Akaike weight of a model can be interpreted as the probability
that the model is the 'best' model among a set of $R$ models. For the $i$th
model, the Akaike weight is given by:
\begin{equation}
w_i=\frac{\mbox{ exp}(-\frac{1}{2}\Delta_i)}{ \sum_{r=1}^R\mbox{
exp}(-\frac{1}{2}\Delta_r)}
\end{equation}
where $\Delta_i$ is defined by
\begin{equation}\label{delta}
\Delta_i=AICc_i-\min(AICc)
\end{equation}
and AICc is a modified version of Akaike's Information Criterion (AIC) that
corrects for small sample sizes relative to the number of parameters. Small values
of AICc are preferred. The AIC approach is a method for comparing the
goodness-of-fit of nested and non-nested models and discourages the
use of models with too many parameters that overfit the data (Burnham and
Anderson, 2002). AICc is defined as
\begin{equation}
AICc=-2\log L+2k\frac{n}{n-k-1}
\end{equation}
where $n$ is the sample size. AICc$_i$ is the AICc value for the $i$th model.

\subsubsection{Basic Reproduction Numbers}
\par Basic reproduction numbers, $R_0$, were also calculated to evaluate the invasive
potential of the disease, based on the model. The basic reproduction number is the average number
of secondary infections produced when one infected individual is introduced into a host population where everyone is susceptible (Anderson and May, 1991). Here, the individual will refer to a herd. When $R_0<1$, the infective may not transmit the disease during the infectious
period and so the disease will die out in the long run (no epidemic). However, if
$R_0>1$, the disease will spread and there will be an epidemic. An epidemic will
occur if  $dI/dt>0$. The derivation of an analytical expression for $R_0$ depends on the epidemiological model. Hence, for the model with density-dependent between-herd
transmission we have:
\begin{align}
\frac{dI}{dt}=U\left(\beta I+k\right)-I c>0\\
\Rightarrow \frac{U\left(\beta I+k\right)}{c}>I \nonumber\\
\Rightarrow \frac{U \beta}{c}+\frac{k}{Ic}>1 \nonumber
\end{align}
and $R_0$ is defined as
\begin{equation} \label{R_0dd}
R_0=\frac{U \beta}{c}+\frac{k}{Ic}
\end{equation}

Similarly, for the model with frequency-dependent between-herd transmission the
basic reproduction number is
\begin{equation} \label{R_0fd}
R_0 = \frac{U \beta}{Nc}+\frac{k}{Ic}
\end{equation}

For the model with no between-herd transmission (i.e. $\beta=0$) the
basic reproduction number is
\begin{equation} \label{R_0_nobeta}
R_0 =\frac{k}{Ic}
\end{equation}

Equation (\ref{R_0_nobeta}) holds for both density- and frequency-dependent
transmission.
The reproduction numbers for the four counties $j=1,\ldots,4$, were calculated by
substituting the parameter estimates of the best fitting model and associated
values of $U_j$ and $N_j$ into the above equations.

\subsubsection{Model Validity}
Green and Medley (2002) argue that proof of the accuracy or validity of any model
should be required before it is used to influence policy. The most important
aspect of this is that a valid model should be true for data not used in the modeling
process. Therefore, the parameter estimates of the best fitting model to data set 1 were
applied to the next year of data
(i.e. badger data from 1998/1999 and cattle data from 1997/1998) to get estimates of the
number of restricted herds, which were then compared to the observed numbers of restricted herds.

\par Models also need to be checked for identifiability, as was done in modeling mastitis in dairy cows by White et al. (2001). In the models presented here, we note that if
there exist two pairs, ($\alpha_1$, $\beta_1$) and ($\alpha_2$, $\beta_2$) that
give the same $I^*$ value then the maximum likelihood estimates of $\alpha$ and
$\beta$ will not be unique. There were four counties so it would be extremely
unlikely to get two $\alpha$ and $\beta$ pairs that give the same  $I^*_j$, for
all $j=1,\ldots,4$. Numerical checks were undertaken, however, to ensure this had
not occurred. The log-likelihood of each model, given by equation (\ref{sumbin}),
was computed for a grid of $\alpha$ and $\beta$ pairs and distinct values were
obtained over the grid to indicate that identifiability was not a problem.

Another issue in model validity is that the model should not be considered
complete until sensitivity analysis is used to identify the rates that have a
large impact on the modeling process. It is important that sensitive rates are
estimated correctly; if the data for these rates are poor, then more data are
required.

A sensitivity analysis similar to that described in Cross and Getz (2006) was carried out to determine the relative importance of several parameters of interest. Eight random values for six parameters of interest were chosen from uniform distributions bounded by the minimum and maximum values, as listed in Table~\ref{sens}. Over 250,000 parameter sets were created from all possible combinations of the eight random values for the six parameters.  For each parameter set, we calculated the total predicted infected herds ($I^*c$) based on the best fitting model to data set 1.

Each of the six parameters were standardised by transforming them into the percentage difference from the mean (i.e.	$(x_i	-\bar{x} )/\bar{x}$ , where $x_i$	is the value of the parameter on run $i$ and $\bar{x}$ is the mean from all runs). A linear regression with the total predicted infected herds as the dependent variable and the six standardised model parameters as the independent variables was carried out to determine which parameters were of significant importance.

In addition, another sensitivity analysis was conducted where parameter sets were
created with $p$ (the length of time on MC), $s$ (SICTT sensitivity) and $a/c$ (the proportion
of infected herds detected via slaughterhouse surveillance) set to their minimum and maximum
likely values as indicated in Table~\ref{sens}. Model fitting of models 1 to 11 was
repeated to determine if model selection and parameter estimates ($\hat{\alpha}$ and $\hat{\beta}$)
remain stable across varying values of the other parameters.

Data analyses were performed in R version 2.12.1.
\section{Results}
\subsection{Results from model fitting}
From the analysis of the proportion of cattle herds with newly detected bTB
infection in a year, the model that best fit data set 1 was model 2, with the
highest Akaike weight of 0.520 (see Table~\ref{results1}). Model 2 assumes no
between-herd transmission ($\beta=0$) and assumes badger-to-herd transmission is
proportional to the total number of infected badgers culled ($\hat{\alpha}=0.0022$, 95\% C.I.: 0.0018-0.0026). Thus, for
example, the total number of infected badgers culled in Cork is 68, therefore, the
rate of infection from wildlife in Cork for that year is
0.0022$\times$68=0.15. Hence, the average length of time before a herd becomes
infected due to wildlife in Cork is 6.68 years (i.e. 1/0.15).
Based on the results from model 2, the percentage of herds expected to experience
a bTB herd breakdown in a year in Cork, Donegal, Kilkenny and Monaghan (i.e.
$\hat{q_j}=I^*_jc$/$N_j$ by equation~(\ref{icnj}) are 12.4\%, 5.5\%, 4.5\% and
5.8\% respectively.
\par Substituting the maximum likelihood estimates of the
unknown parameters from model 2 into $I_j^*c/N_j$, $j$=1,2,3,4
(Equation~(\ref{icnj})) (where $I^*$ for model 2 is defined in
Equation~(\ref{mod2eq}) with $k=\alpha I_w$) and then into equation (\ref{R_0_nobeta}),
gives the reproduction number for the jth county, $j=1,\ldots,4$
\begin{align}
R_0&=\frac{k}{Ic}=\frac{\alpha I_{w j}}{I^*_jc}=\frac{(0.0022)I_{w
j}}{I^*_j(1.25)}
\end{align}

For Cork, Donegal, Kilkenny and Monaghan the reproduction numbers are 0.004,
0.003, 0.005, and 0.002 respectively. All are very
similar, and below 1, hence signifying that bTB will die out in the long-run.
\par Models 6 and 10 both have Akaike weights of 0.229, signifying that there is
substantial support for them also. However, they are essentially the same models
as model 2 since they both have badger-to-herd transmission proportional to the
total number of infected badgers culled ($\hat{\alpha}=0.0022$), with model 6
having density dependent between-herd transmission and model 10 frequency
dependent, but the between-herd transmission coefficient is not significantly
different from 0 in either of the two models. All other models received
essentially no support.
\par
\begin{figure}[H]
\centering
\caption{The four circles represent the observed proportion of herds which
experience a bTB herd breakdown in a single year in the four counties in the
analysis and the solid line represents the fitted model of the proportion of herds
that experience a bTB herd breakdown in a year ($I_j^*c$/$N_j$,
Equation~(\ref{icnj})) as a function of the number of infected badgers
culled in a year.  The data used are from data set 1 (see Table~\ref{datf}). The
parameter estimates are taken from the model that best fits data set 1 based on
Akaike weights, i.e. model 2. }
\centering
 \includegraphics[width=.9\textwidth]{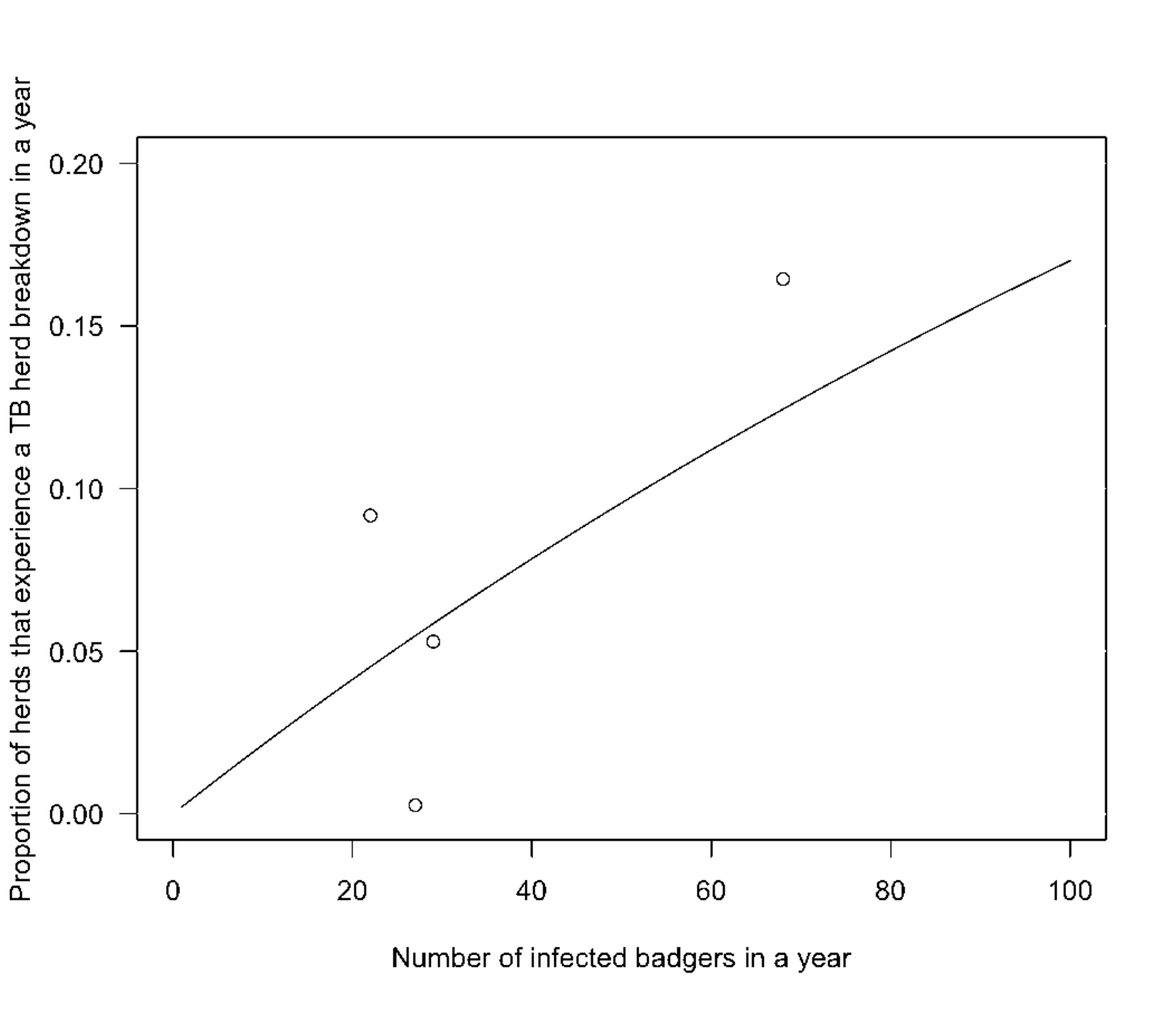}
\label{fitted}
\end{figure}

 A plot of the proportion of herds predicted to experience a bTB herd breakdown in
a single year under model 2, versus the total number of infected badgers culled
($I_w$) is shown in Figure~\ref{fitted} (based on dataset 1). Since $\hat{\alpha} > 0$, Figure 2 shows a positive relationship
between the proportion of newly infected cattle herds in a year and the number of
infected badgers culled in a year. It also shows the implications, based
on the model, that a highly infected badger population would have on the infection rates of the cattle
population, assuming that infection rates of the culled badgers is indicative of the infection rates of the whole badger population. As the infected badger population grows, so too will the
proportion of infected herds in the population.  However, the fitted model is
based on only four empirical observations and that for county Cork is highly
influential in the fitted model.

	Based on the best model fit to data set 1, model 2, a 50\% reduction in the proportion of infected badgers culled would result in a 46\% reduction of bTb incidence in herds in Cork and a 48\% reduction in the other three counties- Donegal, Kilkenny and Monaghan.
\par The results for data sets 2, 3 and 4 are very similar to those of data set 1. The
best fitting model in each case is model 2, where $\beta=0$ and the estimated
values for $\alpha$ differ only slightly from that of data set 1 (data set 2:
$\hat{\alpha}=0.0013,  95$\% $  C.I.: 0.0010-0.0017; $ data set 3:
$\hat{\alpha}=0.0028;  $ 95\% $ C.I.: 0.0024-0.0032;$ data set 4:
$\hat{\alpha}=0.0188;  95$\% $ C.I.: 0.0159-0.0222).$
The results for data set 5 from the pre-study period only are completely different from those obtained for the other four data sets, with model 8 as the best fitting
model, which estimates a non-zero frequency-dependent between-herd transmission
parameter ($\hat{\beta}=2.2473$,  95$\% $  C.I.: 2.0873-2.4073) and no transmission from wildlife ($\alpha=0.0$).

\subsection{Model Validity}
\subsubsection{Prediction}
The parameter estimates of the best fitting model to data set 1 (i.e.
$\hat{\beta}=0$ and $\hat{\alpha}=0.0022$) were applied to the next year of data
(i.e. badger data from 1998/1999 and cattle data from 1997/1998) to get estimates
$I_j^*$, the number of restricted herds.
Values of $I_j^* c$=16, 6, 6, 21 for the four counties respectively were obtained compared to
the observed numbers of restricted herds $29, 3, 14, 19$.
 The lack of agreement reflects the yearly variation in herd bTB incidence in these data and large changes in the badger population due to proactive badger culling.
\par A question also arises as to how critical the assumed badger-to-herd infection
rates are in these models, as model 8 is best for the pre-study period while model
2 is best for the study period. The fifth data set used data on badgers from the
pre-study period only, when few badgers were culled. In all models, it is assumed that the number
of badgers culled is equal to the badger population size and thus the badger
population size is greatly underestimated for data set 5. Models were re-fitted with adjustments to the population size, to examine how results changed. Assuming $N_w$ and $I_w$ as in the first year of the study
period for the data of pre-study, all models were re-fitted to this data set and the best one found
i.e. $N$ and $B$ from data set 5 were used with the $N_w$ and $I_w$ from data set 1.
The results for this adjusted data set 5 were similar to those for data set 1. Model
2 (and 6, 10) was the best fit with no between-herd transmission ($\beta=0$) and
the badger-to-herd transmission rate is proportional to the total number of
infected badgers culled, estimated as ($\hat{\alpha}=0.0164$, 95\% C.I.:
0.0147-0.0182). This is higher than the estimate for data set 1 ($\beta$=0,
$\hat{\alpha}$=0.0022, 95\% C.I.: 0.0018-0.0026).
\subsubsection{Identifiability}
To check the identifiability of the models, the log-likelihood of each model, given by equation (\ref{sumbin}),
was computed for a grid of $\alpha$ and $\beta$ pairs and distinct values were
obtained over the grid indicating identifiability was not a problem.

\subsubsection{Sensitivity analysis}
Table ~\ref{reslts} shows the results of the sensitivity analysis. The table of results ranks parameters in decreasing order of standardised coefficients (i.e. $b/$S.E.) size. The standardised coefficients allow us to compare the relative importance of the six parameters.
The results indicate that all six parameters are important in the model, $N$, the total number of herds, being the most important and herd test sensitivity, $s$, being the least.  It can also be seen that the greater the rate of disease transmission from badgers, the greater the predicted number of infected cattle. The average length of time a herd spends on MC in years also influences the course of the epidemic. The longer the average length of time a herd spends on MC, the smaller the predicted number of infected cattle. Also, the greater the sensitivity of the SICTT herd test, the greater the predicted number of infected cattle, as to be expected.

In the second sensitivity analysis several parameter sets were  created with $p$ (the length of time on MC), $s$ (SICTT sensitivity) and $a/c$ (the proportion
of infected herds detected via slaughterhouse surveillance) set to their minimum and maximum
likely values as indicated in Table~\ref{sens} and models 1 to 11 were re-ran. For all parameter sets, the model selection and parameter estimates ($\hat{\alpha}$ and $\hat{\beta}$) remained unchanged thus indicating the robustness of the results to changes in values of the parameters $p$, $s$ and $a/c$.
\section{Discussion}
\par The study has been performed on a small set of observational data and there
is very little repetition in the data since there were only four counties, thus we cannot draw conclusions from the results with great certainty. However,
the herd population size in each of the four counties studied was large. Thus,
within the limitations above and other study limitations described below,
estimated model parameters are precise.
\par Five data sets were taken from the cattle and badger data from the FAP in Ireland
and for four of them, the model that best fit assumed transmission from
badgers to cattle was proportional to the total
number of infected badgers culled and had no herd-to-herd-transmission. The assumption regarding the badger population size was critical in the
differing model selection for the fifth data set. When this was adjusted and a more realistic badger population size assumed, results similar to the other data sets
were obtained. These findings suggest herd-to-herd transmission is of much
lesser importance for these areas than badger-to-herd transmission.
\par The results indicate that there is a significant association between levels of
bTB in badgers and cattle. This does not prove causality but is, however, in
agreement with results from the FAP and RBCT where some measure of causality was
established (Griffin et al., 2005; Bourne et al., 2007). Reductions in cattle bTB incidence, due to proactive culling of badgers, ranged from 51\% to 68\% over a five-year culling period in the FAP and 23\% in the RBCT. However, since herd bTB was not eliminated, these results also indicated there are sources of infection other than the badger.
\par In this model, it was
assumed that there were only two possible sources of infection of cattle herds,
between-herd infection and infection from wildlife, namely wild badgers, to cattle.
Infection from any other wildlife, such as deer, was ignored, as was within-herd
infection (i.e. cattle-to-cattle infection in a single herd). As in Barlow et al. (1998), re-infection of wildlife, in our case wild badgers, from cattle herds was also considered to be negligible and the
inferences made here were conditional on any such reinfections being negligible.
In addition, it was assumed that cattle infection, as shown by reaction on a skin test, was equivalent to the animal
being infectious and that there is no carrier state in cattle or badgers. Issues in relation to animal testing are discussed in Clegg et al. (2011).
\par In human populations, one-third of the world's population is infected, either
latently or actively, with tuberculosis (Ozcaglar et al., 2012). The rate of latent infection of cattle
herds in Ireland has yet to be established. Ozcaglar et al. (2012) in their
simulation study, showed that a human
tuberculosis epidemic can be viewed as a series of linked subepidemics: a fast
tuberculosis subepidemic driven by direct progression, a slow
tuberculosis subepidemic driven by endogenous reactivation, and
a relapse tuberculosis subepidemic driven by relapse cases; thus,
proving that young and mature tuberculosis epidemics behave differently
and suggested that different control strategies may be necessary for
controlling each subepidemic. Thus, the issue of latent infection of a herd
is an important aspect of bTB epidemiology as this will perhaps drive a
subepidemic by endogenous reactivation.  A similar statement can be made in relation to
wild badgers and has important implications for vaccine testing also, now underway (Aznar et al., 2011). Previous history of infection in a herd has been shown to be a risk factor for bTB in many studies (Griffin et al., 2005; Bourne et al., 2007) and this may be related to latent infection. Thus, the important aspect of epidemic modelling of latent infection with bTB needs study.
\par Epidemic models are particularly suitable for investigating the likelihood of persistence versus fade-out of infection. Blower et al. (1995) demonstrated that it takes several hundred years for a tuberculosis epidemic in humans to rise, fall, and reach an endemic state. Our models estimated approximate reproduction numbers of less than 1, indicating the epidemic would ultimately fade-out, although slowly. This was an asymptotic result and assumed a constant herd population size, while in fact, the herd population size in Ireland is continually changing  (Kelly et al., 2008). More accurate estimates of reproduction numbers and epidemic length would require a separate simulation study. For example, Vynnycky and Fine (1998) include estimates of infection and re-infection rates over time for different ages, and include rates at which individuals who have been infected or reinfected for a time without developing disease move into the `latent' class, in their models.
\par The models used here assumed the population of herds was closed - no immigration i.e. no recruitment. However, from the sensitivity analysis, we can conclude that bTB transmission is influenced by the herd population size i.e. the number of herds. This suggests that increases in population size as a result of economic or other policies may impact strongly on the course of the disease. The assumption of a closed herd population also implies there are no introduced cattle into any herd over the study period and no long range cattle movements. Both of these assumptions are unrealistic, particularly in the Irish context and both factors are known risk factors in herd bTB (Sheridan, 2011).
\par Our results are quite different to those of Donnelly and Hone (2010) that were based on data from the RBCT. They found the model that best fit the RBCT data, using a data set comparable to data set 1 here, was model 11, which has frequency-dependent between-herd transmission
and badger-to-herd transmission proportional to the proportion of infected badgers culled.
 They found much stronger support for frequency-dependent badger-to-cattle
transmission than density-dependent. Our results, based on the  FAP data, suggest that
there is stronger support for density-dependent badger-to-cattle transmission, with
the wildlife transmission variable being the number of infected badgers culled.
In a simple comparison we found that the biggest difference between the RBCT data used by Donnelly and Hone (2010) and the FAP data used here, was that the proportion of infected badgers is much higher in each of the four removal areas in the FAP than in the majority, though not all, badger proactive culling areas of the RBCT. The FAP and RBCT are alike in terms of numbers of herds per km$^2$ and infection rates in herds.

It should also be noted that the models used here are idealised in form and
in addition to the limitations described above, they do not
take into account other features of the epidemic in question. For example, it is assumed the rate of badger-to-cattle herd transmission is independent of herd size and this may not be true. Herd size has been shown in many studies to be one of the most important risk factors for herd bTB (Griffin et al., 2005; Bourne et al., 2007; Kelly et al., 2008) and the sensitivity analysis above showed changes in rates of badger-to-cattle herd transmission has a dramatic effect on model results. Thus, results should be interpreted as representing transmission dynamics with transmission rates averaged over herd sizes.

Farm management practice changes, control policy changes, economic changes and
climate changes may all affect the course of an epidemic and these factors require separate study. In addition spatial correlation structures related to transmission may also be important. Cowled et al. (2012) found that epidemic models for swine fever in wild
pigs in Australia that do not take realistic spatial structures of the wildlife
population into account may overestimate the rate at which a disease will spread
and overestimate the size of an outbreak. A more detailed and complete study could be considered for the future.

However, this study does provide further evidence for the importance of the role of wild badgers on bTB levels in cattle herds in Ireland. This study also demonstrates that, unlike in Great Britain (based on the RBCT), herd-to-herd transmission of bTB is of much lesser importance. \\

\newpage
\begin{table}[H]
\addtolength{\belowcaptionskip}{10pt}
\caption{The five data sets used in the analyses. See Section~\ref{data} for a full description of each data set. }
    \begin{tabular}{| p{2.2cm}  p{2.2cm} p{2.2cm} p{2.2cm}  p{2.2cm} |}
   \hline
    Area & Total Herds\newline (N) & Total Herds Restricted (B)& Total Badgers
    Culled \newline (N$_{\mbox w}$) & Total Infected\newline Badgers Culled
    (I$_{\mbox w}$) \\ \hline\hline
    \multicolumn{5}{|c|}{Data set 1} \\ \hline
    Cork & 292 & 48 & 235 & 68 \\ \hline
    Donegal & 379& 1 & 190 & 27 \\ \hline
    Kilkenny & 229 & 21& 188 & 22 \\ \hline
    Monaghan & 680 & 36 & 165 & 29 \\ \hline\hline

    \multicolumn{5}{|c|}{Data set 2} \\ \hline
    Cork & 288 & 29 & 235 & 68 \\ \hline
    Donegal & 375& 3 & 190 & 27 \\ \hline
    Kilkenny & 230 & 14& 188 & 22 \\ \hline
    Monaghan & 687 & 19 & 165 & 29 \\ \hline\hline

    \multicolumn{5}{|c|}{Data set 3} \\ \hline
    Cork & 288 & 67 & 446 & 115 \\ \hline
    Donegal & 375& 14 & 273 & 40 \\ \hline
    Kilkenny & 230 & 34& 409 & 56 \\ \hline
    Monaghan & 701 & 112 & 414 & 66 \\ \hline\hline

    \multicolumn{5}{|c|}{Data set 4} \\ \hline
    Cork & 294 & 230 & 467 & 116 \\ \hline
    Donegal & 396 & 87 & 282 & 44 \\ \hline
    Kilkenny & 233 & 119 & 618 & 107 \\ \hline
    Monaghan & 701 & 338 & 627 & 88 \\ \hline\hline

    \multicolumn{5}{|c|}{Data set 5} \\ \hline
    Cork & 294 & 163 & 21 & 1 \\ \hline
    Donegal & 396 & 73 & 9 & 4 \\ \hline
    Kilkenny & 233 & 85 & 209 & 51 \\ \hline
    Monaghan & 680 & 226 & 213 & 22 \\ \hline

    \end{tabular}
    \label{datf}
\end{table}

\newpage

\begin{table}[H]
\addtolength{\belowcaptionskip}{10pt}
\caption{Sensitivity Analysis - Parameter Boundaries}
    \begin{tabular}{| p{5cm}  p{1.1cm}  p{1.5cm}  p{1.6cm}  p{3cm} |}
   \hline
    Parameter & Symbol  & Minimum & Maximum & Source \\ \hline\hline
    Total number of herds & $N$ & 200 & 1,000 & From data in this paper \\ \hline
    Average length of time a herd spends on MC in years & $p$& 0.329 & 0.7 & M. Good, DAFM, personal communication \& Donnelly and Hone (2010) \\ \hline
    Herd test sensitivity rate in per cent& $s$ & 0.529& 0.606 & Clegg et al. (2011)\\ \hline
    Rate of infected cattle detection through slaughterhouse surveillance per annum& $a$ & 0.27 & 0.46 & Frankena et al. (2007)\\ \hline
    Proportionality constant associated with the rate of infection from wildlife  & $\alpha$ & 0 & 0.8 & From data in this paper\\ \hline
    Total number of infected badgers culled & $I_w$ & 10 & 100 & From data in this paper \\ \hline
    \end{tabular}
    \label{sens}
\end{table}

\newpage

\begin{table}[H]
\begin{threeparttable}[b]
\addtolength{\belowcaptionskip}{10pt}
\caption{Linear regression sensitivity analysis using over 250,000 runs of model 2 with parameter values chosen from uniform distributions and total predicted infected herds as the dependent variable.}
    \begin{tabular}{| p{5cm}  p{1.3cm}  p{2cm} p{2cm}  p{2cm} |}
   \hline
    Parameter\tnote{1}& Symbol  & Coefficient ($b$) \tnote{2}  & Standard Error (S.E.) & $b/$S.E. \\ \hline\hline
    Total number of herds & $N$ & 412.26 & 0.39 & 1051.68  \\ \hline
    Proportionality constant associated with the rate of infection from wildlife  & $\alpha$ & 95.51& 0.23 & 412.92 \\ \hline
    Rate of infected cattle detection through slaughterhouse surveillance per annum& $a$ & 126.74& 0.95 & 133.75\\ \hline
    Total number of infected badgers culled & $I_w$ &  32.76 & 0.27 & 119.96 \\ \hline
     Average length of time a herd spends on MC in years & $p$& -143.01 & 1.49 & -95.75\\ \hline
    Herd test sensitivity rate in per cent & $s$ & 317.2 & 4.34 & 73.05\\ \hline
    \end{tabular}
    \label{reslts}
    \begin{tablenotes}
      \item[1] \footnotesize{All parameters were  transformed to percentage difference from the mean (i.e.	$(x_i	-\bar{x} )/\bar{x}$ , where $x_i$	is the value of the parameter on run $i$ and $\bar{x}$ is the mean from all runs).}\\
      \item[2] \footnotesize{All coefficients have a p-value less than 0.001. }
    \end{tablenotes}
    \end{threeparttable}
\end{table}

\newpage

\begin{landscape}
\ctable[doinside=\normalsize, caption={Parameter estimates and log-likelihood values from the eleven
varying models fitted to data set  1 (see Table~\ref{datf}) on bovine tuberculosis in cattle and
badgers from the Four Area Project. Each of the models represents some combination of the two fitted parameters $\beta$
and $\alpha$, where $\beta$ is a measure of herd-to-herd transmission per annum (Density Dependent (DD) or Frequency Dependent (FD)) and $k$ is a measure of badger-to-herd transmission per annum, such that $k=\alpha N_w$, $k=\alpha I_w$ or $k =\alpha(I_w/N_w)$, where $N_w$ equals the number of badgers culled in an area and $I_w$ equals the number of infected badgers culled in an area. Model 2, highlighted in dark grey, has the
most support from the data with an Akaike weight of 0.52.},label=results1,pos=hbp!]{|p{1cm} p{2cm} p{2.1cm} p{1.8cm} c p{1.3cm} c p{1.3cm} p{1.6cm}  p{1.1cm} p{1.1cm}|}{}{
\FL
Model&Between-herd\newline transmission  & Transmission from \newline wildlife\footnotemark[1] &
Number \newline of \newline parameters & $\beta$ & p-value  $H_0:\newline \beta=0$
& $\alpha$ & p-value\newline $H_0:\newline\alpha=0$ & Log-\newline likelihood &
AICc\footnotemark[2] & Akaike \newline weight \NN \hline\hline
1&None & $\propto N_w$ & 1 &- -\footnotemark[3] &- -\footnotemark[3]  & 0.0004 &
N/A\footnotemark[4]  & -382.4703 & 766.94 & 0.000\\ \hline
\rowcolor{black!35} 2&None & $\propto I_w$ & 1 &  - -\footnotemark[3] & -
-\footnotemark[3] & 0.0022 & N/A\footnotemark[4]  & -372.4763 & 746.96
&0.520 \\ \hline
3&None & $\propto I_w/N_w$ & 1 & - -\footnotemark[3]  & - -\footnotemark[3]  &
0.4162 & N/A\footnotemark[4]  & -376.2648 & 754.53 & 0.012 \\ \hline
4&DD & None& 1 & 0.0006 & N/A\footnotemark[4]  &- -\footnotemark[3]  & -
-\footnotemark[3]  & -752.6828 & 1507.39 & 0.000 \\ \hline
5&DD  & $\propto N_w$ & 2 & 0.0000 &0.942  & 0.0004 &  $<$0.001  & -380.3202 &
764.65 & 0.000 \\ \hline
\rowcolor{black!17}6&DD& $\propto I_w$ & 2 & 0.0000 &0.968 & 0.0022 & $<$0.001
& -372.2951 & 748.60 &0.229 \\ \hline
7&DD & $\propto I_w/N_w$ & 2 & 0.0000 & 0.978& 0.4098 &$<$0.001  & -376.1010 &
756.21 & 0.005 \\ \hline
8&FD & None & 1 & 1.3703 & N/A\footnotemark[4]  &- -\footnotemark[3]  & -
-\footnotemark[3] & -388.7466 & 779.50 & 0.000 \\ \hline
9&FD & $\propto N_w$ & 2 & 0.0000 & 0.999& 0.0004 &$<$0.001 & -382.3800 & 768.77
&0.000 \\ \hline
\rowcolor{black!17}10&FD & $\propto I_w$ & 2 & 0.0000 & 0.993& 0.0022 & $<$0.001
& -372.2950 & 748.60 &0.229\\ \hline
11&FD & $\propto I_w/N_w$ & 2 & 0.0000 & 0.994 & 0.4101 & $<$0.001 & -376.1002 &
756.21& 0.005 \NN \hline
}
\setlength{\footnotesep}{0.45\baselineskip}
\footnotetext[1]{$\propto$: proportional to.}
\footnotetext[2]{AICc: Akaike Information Criterion corrected for small sample sizes
relative to the number of parameters.}
\footnotetext[3]{In the case when we assume one of the transmission parameters,
$\beta$ or $\alpha$, is zero, we omit the parameters estimate and no p-value is
calculable.}
\footnotetext[4]{When we assume one of the transmission parameters, $\beta$ or
$\alpha$, is zero, the calculation of the p-value of the other transmission
parameter is not applicable (N/A) since the null hypothesis would state that both
parameters were zero and hence, there would be no transmission of tuberculosis to
cattle.}

\end{landscape}

\newpage

\end{document}